\documentclass[a4paper]{article}
\usepackage[T1]{fontenc}
\usepackage{authblk}
\usepackage{graphicx}

\title{CAST microbulk micromegas \\ in the Canfranc Underground Laboratory}
\author[1]{A.~Tom\'as}
\author[2]{S.~Aune}
\author[1]{T.~Dafni}
\author[3]{G.~Fanourakis}
\author[2]{E.~Ferrer-Ribas}
\author[1,2]{J.~Gal\'an}
\author[1]{J.A.~Garc\'ia}
\author[patras]{A.~Gardikiotis}
\author[4]{T.~Geralis}
\author[2]{I.~Giomataris}
\author[1]{H.~G\'omez}
\author[1]{J.G.~Garza}
\author[1]{D.C.~Herrera}
\author[2]{F.J.~Iguaz}
\author[1]{I.G.~Irastorza}
\author[1]{G.~Luz\'on}
\author[2]{T.~Papaevangelou}
\author[1]{A.~Rodr\'iguez}
\author[5]{J.~Ruz}
\author[1]{L.~Segu\'i}
\author[6]{T.~Vafeiadis}
\author[7,8]{S.C.~Yildiz}

\affil[1]{Grupo de F\'isica Nuclear y Astropart\'iculas, University of Zaragoza, Zaragoza, Spain}
\affil[2]{IRFU, Centre d'\'Etudes de 2, CEA, Gif-sur-Yvette, France}
\affil[3]{Institute of Nuclear Physics, NCSR Demokritos, Athens, Greece}
\affil[4]{University of Patras, Patras, Greece}
\affil[5]{CERN, European Organization for Particle Physics and Nuclear Research}
\affil[6]{Aristotle University of Thessaloniki, Thessaloniki, Greece}
\affil[7]{Do\u{g}u\c{s} University, Istanbul, Turkey}
\affil[8]{Bo\u{g}azi\c{c}i University, Istanbul, Turkey}

\begin{document}

\maketitle

\begin{abstract}
During the last taking data campaigns of the CAST experiment, the micromegas detectors have achieved background levels of $\approx 5 \times 10^{-6}$keV$^{-1}$cm$^{-2}$s$^{-1}$ between 2 and 9 keV. This performance has been possible thanks to the introduction of the microbulk technology, the implementation of a shielding and the development of discrimination algorithms. It has motivated new studies towards a deeper understanding of CAST detectors background. One of the working lines includes the construction of a replica of the set-up used in CAST by micromegas detectors and its installation in the Canfranc Underground Laboratory. Thanks to the comparison between the performance of the detectors underground and at surface, shielding upgrades, etc, different contributions to the detectors background have been evaluated. In particular, an upper limit $< 2 \times 10^{-7}$keV$^{-1}$cm$^{-2}$s$^{-1}$ for the intrinsic background of the detector has been obtained. This work means a first evaluation of the potential of the newest micromegas technology in an underground laboratory, the most suitable environment for Rare Event Searches.
\end{abstract}

\section{Introduction}
\subsection{CAST}

\medskip
The principle of an axion-helioscope\cite{Sikivie} consists of the conversion into photons of some of the axions\cite{axions1}\cite{axions2} generated and comming from the Sun in presence of a transverse magnetic field. The solar axion flux could then be identified as an excess of photons registered by X-ray detectors (axion's spectrum is ranges from 2 to 9 keV) during the time the helioscope is aligned with the Sun.  

\medskip
The CERN Axion Solar Telescope, CAST\cite{cast}\cite{exrs}, is the best realization until now of an axion-helioscope thanks to the powerful LHC dipole prototype and to the sensitivity of its X-rays detectors. The typical background rate during current operation phase in CAST\cite{newcast} has been $5 \times 10^{-6}$keV$^{-1}$cm$^{-2}$s$^{-1}$ from 2 to 9 keV. The goal of improving this performance towards a new generation of axion helioscopes\cite{NGAH}, needs a carefully characterization of present detectors. 

\subsection{CAST micromegas detectors}

\medskip

There are three main requirements for low background detectors: the use of low radicative materials, a shielding againts environmental radiation and good discrimination capabilities. The successive introduction and continuous improvement of these strategies has led to a constant decrease of the background level in CAST micromegas detectors.
 
\medskip
In the new generation of micromegas\cite{micromegas}, called microbulk\cite{microbulk}\cite{Paco_microbulk}, all the readout is contained in a 80 $\mu$m foil. This foil has been measured showing good radiopurity\cite{paquito}. The rest of the mechanical structure of the detector is made of Plexiglass (also radiopure) with the exception of the drift cathode which is made of steel. A thin window (only 4 $\mu$m of aluminized mylar) and the operation at slighty overpressure (1.35 bar) optimizes the detector efficiency for X-rays detection. 

\medskip
The detector is coverd by a shielding composed of a first 5 mm thickness copper layer and second one of archeological lead and 25 mm thickness. Also N$_2$ flow is used to avoid Rn emanations nearby the detector. Moreover the readout design allows a descriptive characterization of the registered events, providing both spatial and time features. This information is used to discriminate X-rays events from other interactions\cite{analysis}\cite[Chap. 4 and 5]{Javis}.   

\begin{figure}[!h]
    \centering
     \includegraphics[width = .95\textwidth, angle = 0]{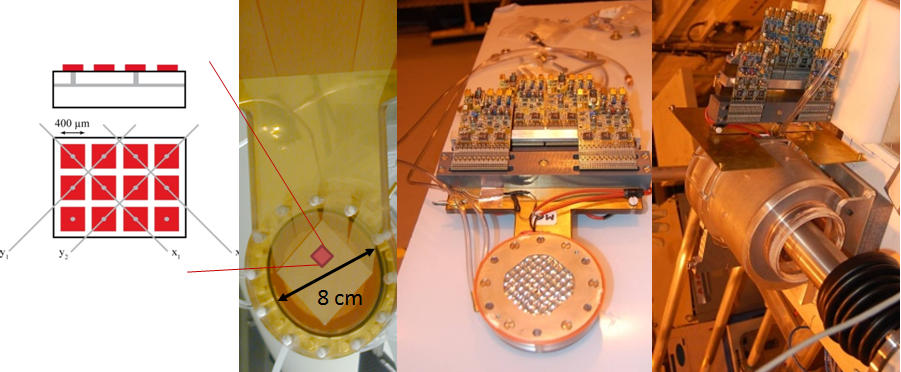}
      \caption{From left to right: pixelized strip readout; micromegas foil (containing amplification structure plus readout plane) glued on a PCB racket; addition of the chamber with a thin aluminized window and the electronic cards; the detector connection to the magnet pipe during an intermediate installation where lead shielding can be observed.}
    \label{fig:castdetectors}
\end{figure}

\subsection{Work motivation and methodology}

A new set-up (Figure 2), that essentially reproduces the CAST one (as described in the previous section), was built in the University of Zaragoza. It is completed with a CAST microbulk detector; the experiments acquisition is also replicated and the same analysis programs are used to process the data. The set-up was equiped with a calibrator containing a $^{55}$Fe source in order to check periodically the detector status.

\begin{figure}[!h]
    \centering
     \includegraphics[width = .48\textwidth]{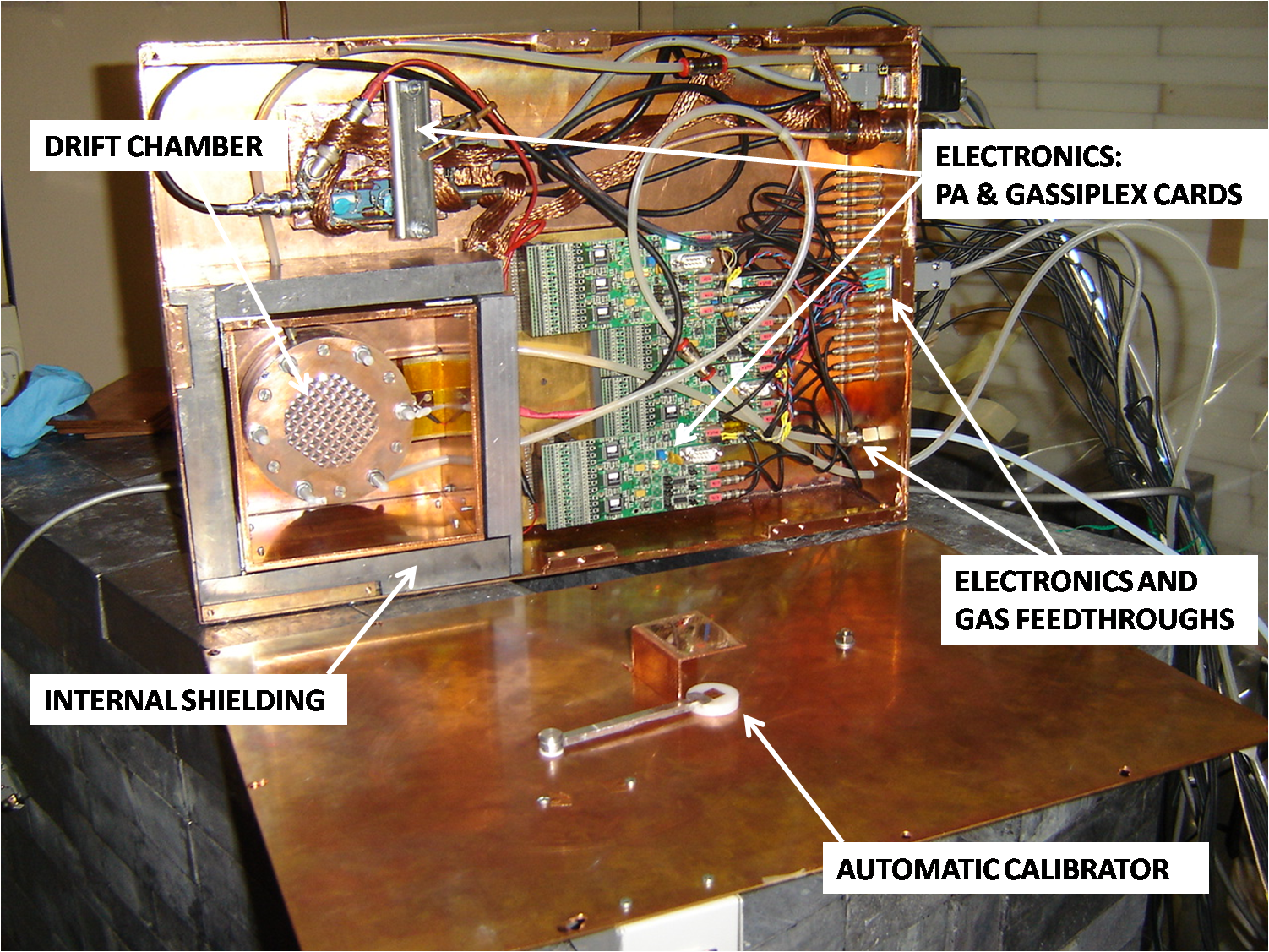}
     \includegraphics[width = .48\textwidth]{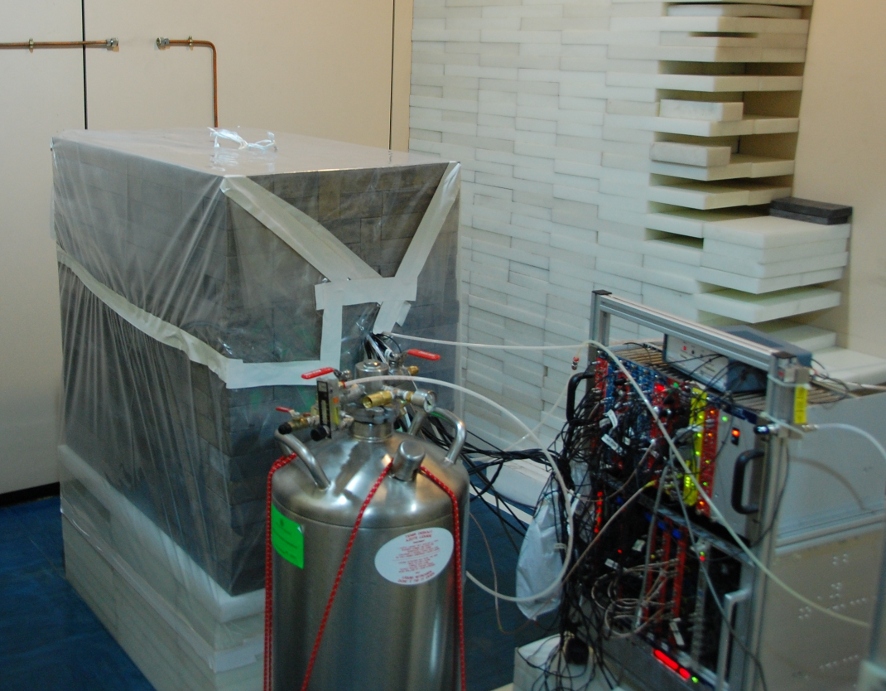}
      \caption{Left: CAST-like test set-up first used in Zaragoza's test bench and later in LSC. Light shielding composed by 2.5 cm of lead with an internal shielding of 0.5 cm of more radiopure copper. The two last lead caps are added after closing the Faraday cage. The set-up is prepared for 6 keV X-Rays automatic calibrations thanks to a $^{55}$Fe source inside the Faraday cage. Right: the same set-up after the later upgrading: a 20 cm thick lead shielding. The liquid nitrogen dewar and the acquisition electronics are shown too.}
    \label{fig:LSCset-up}
\end{figure}

\medskip
The goal of the work is to study the influence of changes in the detector environment on the detector background. For each change a concrete origin of background is pretended to be avoided or substantially reduced. Comparisons between different set-ups and detectors were made too, trying to itemize the several causes which are susceptible to have an influence and evaluate its strength.

\medskip
Moving the detector to the Canfranc Underground Laboratory, at 2500 m.w.e depth, the cosmic muons are reduced by a factor $\times 10^{4}$\cite{LSCmuons}. In contrast, the radon concentration is much more intense than at surface and highly variable, with typical values from 100 to 200 Bq/m$^3$. The absence of cosmic rays allows the use of a much thicker shielding, which practically blocks the environmental radiation.

\section{Surface and underground comparisons}

\medskip

\begin{figure}[!h]
    \centering
     \includegraphics[width = .48\textwidth, angle = 0]{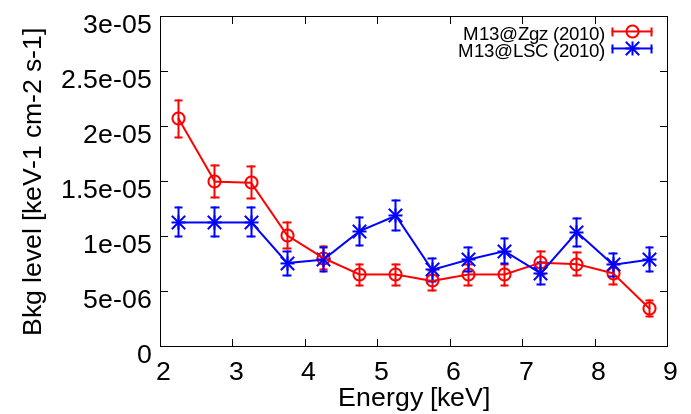}
     \includegraphics[width = .48\textwidth, angle = 0]{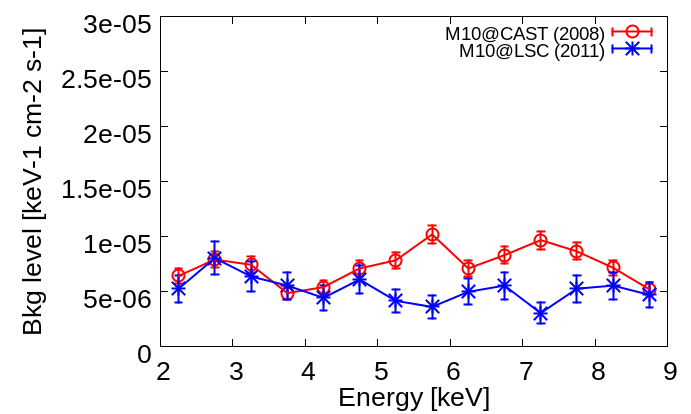}
      \caption{Left: final background spectra taken in identical conditions in the Zaragoza's test-bench (surface) and in the LSC (2500 m.w.e. deep) with the M13 detector.Right: final background spectra taken with the M10 detector in the CAST experiment (sunrise side, Autumn 2008) and in the LSC (Winter 2011) in similar shielding configurations.}
    \label{fig:CASTvsLSC}
\end{figure}

In a first comparison the same detector (Figure 3, left) and set-up (Figure 2, left) are conseved and the environment is changed from 200 meters above the sea level to 2500 m.w.e depth in the Canfranc Underground Laboratory. The trigger rate at Canfranc was $\approx$ 0.2 Hz, about 5 times lower than at surface; however, after application of offline cuts, the final background remained very similar $\approx 5 \times 10^{-6}$keV$^{-1}$cm$^{-2}$s$^{-1}$ in the 2-9 keV energy range(Figure \ref{fig:CASTvsLSC} left). Therefore the cosmic rays cause $\approx$ 80\% of the events registered in the detector in the surface; but the final background, which is obtained after the application of discrimination algorithms, is practically not altered by them.

\medskip
In a second comparison (Figure 3, right) the same detector is used at surface and in Canfranc underground; though inside similar, but not identical, set-ups. First in the CAST experiment, during the 2008 data taking, and later with our test set-up (Figure 2, left). The main difference is the presence of two fluoresce peaks in the CAST spectrum. The first peak is really composed by the fluorescences related with the steel components (Cr, Fe, Ni at 5.4, 6.4 and 7.0 keV). It is related with the stainess steel cathode and the vacuum pipe that connect the detector to the superconductor magnet. The second peak, at 8.0 keV, is a copper fluorescence orginated in the micromegas. None of the spectra taken with the test set-up represented in Figure 3 presents these peaks. The intensification of the fluorescences must be due to the shielding opening in CAST set-up, which is needed to connect to the detector and the magnet, as it was mentioned before. This picture has been confirmed by Monte Carlo simulations of the CAST set-up and, in addition, by the important suppression recently observed in the fluorescence peaks with the micromegas currently taking data by means of an improvement of the front-shielding near the opening region.

\medskip
Finally, a third comparison can be made between the spectra taken by two different detectors in identical conditions: test set-up in Canfranc underground. The first detector (Figure 3, left) was a CAST spare detector which suffered from some physical deffects. In contrast, the second one (Figure 3, right) was at service in CAST obtaining a notable improvement in the final background compared with previous detectors\cite[Chap. 7 Sec. 5]{Javis}. The fact that the background level obtained with the best detector is appreciably lower means that the status of the micromegas, which preserves its discrimination capabilities, is a relevant factor. However, both levels are not very far one to the other, which may stands up to the remained background to be due to real X-rays, and so inavoidable with this set-up.  

\medskip
Regarding the influence of radon in the final background level, it can be said that it is not a limiting factor with this set-up. Even when the radon flow was interrupted, there was not a clear effect.

\medskip
The stainless steel cathode, in principle the unique compound of the micromegas chamber which is not particularly radiopure, was also rejected to have a relevant contribution, at this background level, as not changed was reported by the replacement of this by a replica which was built in copper.

\section{Shielding upgrade and internal configurations}

\medskip

The CAST-like set-up (Figure 2, left) was surrounded with a 20 cm thick of lead extra layer (Figure 2, right). The idea is to evaluate the relative weights from internal contamination and external radiation to the CAST background. Even when cosmic muons can be directly rejected by the CAST detector, the same does not have to be true for the secondary particles produced by the muon interactions, for instance, in the shielding. Because of this reason such a thick shielding may produce a secundary effect at surface. In order to clearly separate different contributions, this test should be done first underground.

\medskip
The shielding upgrade drove to an important reduction of both the trigger rate, by a factor $\approx$40, and the final background by a factor $\approx$30. This fact stands for the interactions in the chamber to have almost exclusively an external origin and stablishes an upper limit for the micromegas detector intrinsic background below $2 \times 10^{-7}$keV$^{-1}$cm$^{-2}$s$^{-1}$ in the 2-9 keV range. It must be noted that we consider the detector as the sum of the microbulk and the chamber structure with a copper cathode.

\medskip
This background could still be influenced by the radioactivity of the shielding materials. The most internal layer of the shielding is made or radiopure copper, thought inside the Faraday cage (Figure 3) the detector internal cavity is not perfectly closed, as a gap is needed by the calibrator to access to the detector. Moreover, it must be remarked that no archeological lead has been used for the shielding in the test set-up (in contrast with actual CAST set-ups). With aim of clarifying this a new setting was tried: the chamber window was closed with a 1.5 mm layer of copper, just leaving a small hole for calibrations. If an important contribution to the background were due to soft X-rays comming from the shielding walls, they could access the chamber volume only trought the transparent chamber window, which has been blocked in the second setting. No apprecible change was detected in the background level and shape (Figure 5 and Table 1).

\begin{figure}[!h]
    \centering
     \includegraphics[width = .48\textwidth, angle = 0]{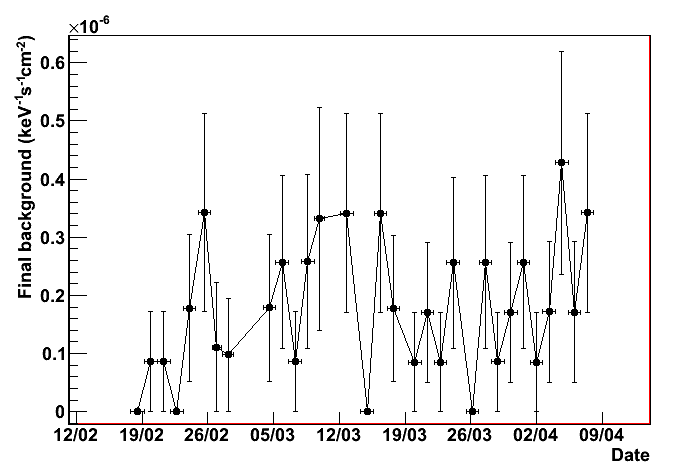}
	\includegraphics[width = .48\textwidth, angle = 0]{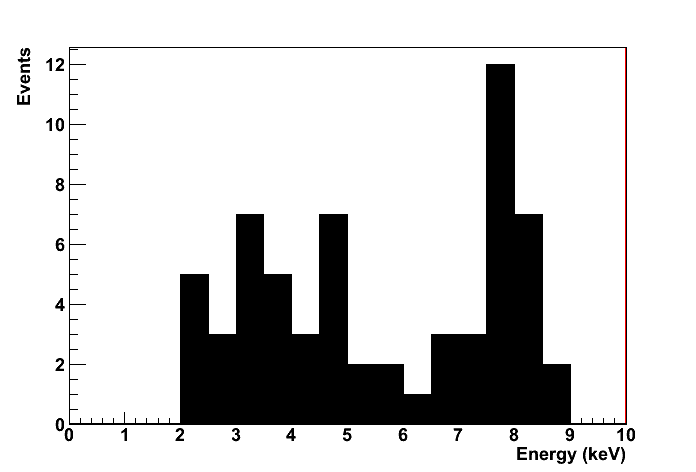}
      \caption{Left: Final background evolution after the shielding upgrade. Right: final background spectrum.}
    \label{fig:ULBPgeneral}
\end{figure}

\begin{figure}[!h]
\begin{center}
	\begin{tabular}{l|ccc}
		 & Window closed & Window open & All the period \\ \hline
		2-7 keV & $1.5 \pm0.9$ & $1.6 \pm 0.9$ & $1.5 \pm 0.6$ \\
		2-9 keV & $1.8 \pm 0.8$ & $1.7 \pm 0.8$ & $1.7 \pm 0.3$ \\
	\end{tabular}
	\label{tab:ULBPdetailled}
\end{center}
\begin{center}
Table 1: Final background levels expressed in $10^{-7}$keV$^{-1}$cm$^{-2}$s$^{-1}$.
\end{center}
\end{figure}

To sum up, a background level of about 1 count/day (30 times lower than the nominal value in CAST) was stable during  41.4 days of live time. A final amount of 41 counts conformed a spectrum with a small, but clear, copper fluorescence peak. 

\section{Summary and conclusions}

\medskip
LSC is a useful test bench for CAST detectors as it allows us to analyze the relative weight of the different physical contributions to CAST micromegas background. A direct comparison of the background by using the same set-up at surface and underground shows that, even when cosmic rays dominate the trigger rate at surface, muons are efficiently rejected by the micromegas. An important part of the CAST background, in special the fluorescence peaks, is related with the intrusion of gammas from the connection between the detector and the magnet bores. This is already being reduced by means of upgrading the shielding of the line. 

\medskip
CAST micromegas final background level is dominated by the environmental gamma flux, as it is clear from the reduction obtained by shielding upgrade: about 30 times, which leads to an upper limit for the micromegas detector intrinsic background below $2 \times 10^{-7}$keV$^{-1}$cm$^{-2}$s$^{-1}$ in the 2-9 keV range. The work is ongoing to identify more concrete origins for the internal contaminations.

\begin{figure}[!h]
    \centering
     \includegraphics[width = .43\textwidth, angle = 0]{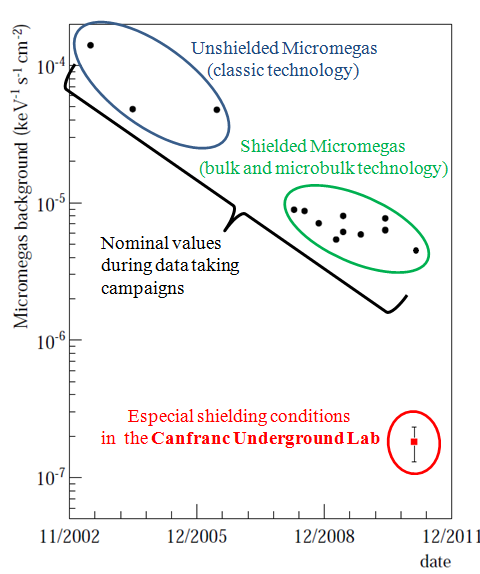}
      \caption{Historical background levels for CAST micromegas detectors.}
    \label{fig:history}
\end{figure} 

\medskip
From this work a new point is added to the CAST micromegas background historical curve (Figure 5). The new point goes away the general tendency, roughly exponential, because it does not correspond to the context of the true CAST experiment. It must be considered as a limit for the future evolution of CAST micromegas background imposed by the intrinsic radioactivity. Although new techniques may bring more radiopure micromegas. It can also represent a first indication for micromegas in a new context: the Underground Physics.

\section{Acknowledgements.}
\medskip
We acknowledge support from the European Commission under the European Research Council T-REX Starting Grant ref. ERC-2009-StG-240054 of the IDEAS program of the 7th EU Framework Program, as well as, from the Spanish Ministry of Science and Innovation (MICINN) under contract ref. FPA2008-03456, and the CPAN project ref. CSD2007-00042 from the Consolider-Ingenio 2010 program of the MICINN. Part of these grants are funded by the European Regional Development Fund (ERDF/FEDER).


\end{document}